\documentclass[acmsmall]{acmart}
\usepackage{graphicx}
\usepackage{color}
\usepackage{todonotes}

\usepackage{algorithm}
\usepackage{algpseudocode}  
\usepackage{siunitx} 
\usepackage{multirow}
\usepackage{booktabs}
\usepackage{subfigure}
\usepackage{hyperref}

\AtBeginDocument{%
  \providecommand\BibTeX{{%
    \normalfont B\kern-0.5em{\scshape i\kern-0.25em b}\kern-0.8em\TeX}}}

\setcopyright{acmcopyright}
\copyrightyear{2021}
\acmYear{2021}
\acmDOI{}

\acmConference[Suzhou]{2021 6th International Conference on Biomedical Signal and Image Processing}{August 20-22, 2021}{Suzhou, China}
\acmBooktitle{2021 6th International Conference on Biomedical Signal and Image Processing, August 20-22, 2021, Suzhou, China}
\acmPrice{}
\acmISBN{978-1-4503-9050-7}

\begin{document}

\title{A Rigid Registration Method in TEVAR}

\author{Meng Li}
\email{122019000311@ccmu.edu.cn}
\orcid{0000-0001-8810-1719}
\author{Changyan Lin}
\authornote{Corresponding author}
\email{llbl@sina.com}
\affiliation{%
  \institution{Beijing Anzhen Hospital, Capital Medical University; Beijing Key Laboratory of Fundamental Research on Biomechanics in Clinical Application; Beijing Institute of Heart, Lung \& Blood Vessel Diseases;}
  \streetaddress{Anzhen Road 2th, Chaoyang District}
  \city{Beijing}
  \state{Beijing}
  \country{China}
  \postcode{100069}
}
  
\author{Heng Wu}
\affiliation{%
  \institution{Beijing Anzhen Hospital, Capital Medical University}
  \streetaddress{Anzhen Road 2th, Chaoyang District}
  \city{Beijing}
  \country{China}}

\author{Jiasong Li}
\affiliation{%
  \institution{Beijing Anzhen Hospital, Capital Medical University}
  \streetaddress{Anzhen Road 2th, Chaoyang District}
  \city{Beijing}
  \country{China}}

\author{Hongshuai Cao}
\affiliation{%
  \institution{Beijing Anzhen Hospital, Capital Medical University}
  \streetaddress{Anzhen Road 2th, Chaoyang District}
  \city{Beijing}
  \country{China}}

\renewcommand{\shortauthors}{Meng Li and Changyan Lin}

\begin{abstract}

Since the mapping relationship between definitized intra-interventional X-ray and undefined pre-interventional Computed Tomography(CT) is uncertain, auxiliary positioning devices or body markers, such as medical implants, are commonly used to determine this relationship. However, such approaches can not be widely used in clinical due to the complex realities. 
To determine the mapping relationship, and achieve a initializtion post estimation of human body without auxiliary equipment or markers, proposed method applies image segmentation and deep feature matching to directly match the X-ray and CT images. As a result, the well-trained network can directly predict the spatial correspondence between arbitrary X-ray and CT. The experimental results show that when combining our approach with the conventional approach, the achieved accuracy and speed can meet the basic clinical intervention needs, and it provides a new direction for intra-interventional registration. 
\end{abstract}




\maketitle

\section{Introduction}
\label{Sec: Intro}

Thoracic Endovascular Aortic Repair (TEVAR) is the preferred treatment for thoracic aortic dilatation diseases as it allows the low mortality and complications \cite{rizvi2009effect}. 
Interventional therapy is often performed on two-dimensional(2D) medical images such as X-ray. 
Due to the lack of clear development of vessels under X-ray fluoroscopy, contrast agents which often cause serious complications have to be used in interventional therapy. 
This paper presents a medical image registration approach based on transformer to accurately develop the aorta and its branches directly on X-ray fluoroscopy. 
Due to the position of aortic arch is fixed by the arterial ligament and three arterial branches, it is considered immobile in proposed approach. 
The feasibility of this idea will be validated in Section.\ref{Sec.feasibility}. 

Machine learning-based medical image registration methods are mostly divided into two categories: optimization-based and learning-based approaches. 
Optimization-based conventional approaches project the 3D CT data to 2D virtual X-ray, according to a series of projection parameters. Then searching for the most similar image to the real X-ray iterativly and determining its spatial transformation parameters.
which is difficult to meet the requirement of real-time registration suffer from extremely high computational cost and sensition of initial estimates\cite{matl2017vascular}. 
Although the recent development of deep learning have been successfully applied to various medical applications \cite{song2018human,ronneberger2015u,ozturk2020automated,song2020spectral}, such techniques have been barely used for 3D CT data to 2D virtual X-ray image registration. Even some works simply applied Convolutional Neural Networks (CNNs), autoencoder (AEs) or other network structures to conduct feature extraction\cite{wu2015scalable,zhao2015deep}or similarity measurement \cite{ghosal2017deep,cheng2018deep}, which direct predict the registration parameter, they haven't solve the problem of time-consuming of iteratively optimization. More importantly, considering the certainty of intra-interventional human position and the uncertainty of pre-interventional CT data acquisition position, researchers had to choose positioning from two perspectives\cite{liao2019multiview:,miao2018dilated} or using traumatic markers like medical implants\cite{miao2018dilated} to solve the problems. In other words, these methods are not widely used in clinic. 

Since pre-interventional 3D CT can provide accurate anatomical information, this paper adopts 2D-3D image registration method, 
The aim is to establish a direct connection between 2D X-ray and 3D CT. The main ideas of proposed method are as follows: two segmentation networks been used to segment skeletal features from two dimension. Then the extracted features are put into the crossmodal matching transformer module to matching features of different dimensions. After the process of learning mapping relationships, we embed five positioning parameters available in interventional imaging equipment into our network to improved the ability of the network. The main contributions of our research are:

\begin{itemize}
\item Application of medical image registration technology in the field of TEVAR interventional therapy.

\item The proposed approach can directly match 2D and 3D images whose latent features. This provide a solution for the precise localization in TEVAR under X-ray, and can potentially reduce the use of contrast agents and the exposure time under radiation.
\end{itemize}

\section{Related Works}
\subsection{Learning-based medical image segmentation}

Although image segmentation technology has been widely used in the field of medicine, the methods used in clinic are still limited. Medical images have abundant spatial information (such as complex texture structure), and the process of network downsampling is easy to lose these spatial information. Encoder-decoder networks with symmetric structures\cite{ronneberger2015u,quan2016fusionnet}. can be better preserve these spatial information. 
Using the same method, Cicek et al. replaced the two-dimensional convolution layer with the three-dimensional convolution layer to construct 3D U-Net, and realized the end-to-end processing of 3D CT\cite{cciccek20163d}.  Although deep learning have shown significant progress compared with conventional algorithms, the process of medical image tagging is time-consuming and labor-consuming, which limits the further development of deep learning algorithms in clinic. Researchers consider more diverse ways of using unlabeled data\cite{xu2016neuron,yi2019generative}, but such segmentation methods are not suitable for registration tasks that are extremely sensitive to spatial position.

\subsection{Multimodal data fusion}

It is a natural idea to fuse the same features extracted from two- and three-dimensional neural networks. S-PCNN\cite{yang2015block} divides the source image into several blocks, then calculate the spatial frequency (SF) of the blocks as linking strength beta of the PCNN. It is used to extract the medical imageures find the best oscillation frequency graph (OFG) iteratively. Although combined the information of multi-modal images, the time-consuming of divided blocks and iteration is not allowed in real-time medical image registration. Parallel cross CNN (PCCNN) model\cite{tang2016parallel} extracts two group features of CNN in parallel through a couple of deep CNN data transform flows. The information of two streams are fused together after the first fully connected layers in each stream. After the final Softmax regression, the 1024-dimensional image feature vector is classified and recognized. The method which introduced depthwise separable convolution combined with depthwise and pointwise convolution to replace the standard convolution as a basic convolution module\cite{li2020deep}, respectively applies the convolution in different channels, and applies a $1\times1$ convolution to combine separate features generated by the depthwise convolution to reduce the complexity and computational cost of model.
The method of fussing information of unaligned multimodal language sequences gives us enlightenment\cite{tsai2019multimodal}. By the method of multimodal transformer, they address the above issues in an end-to-end manner without explicitly aligning the audio, language, vision.

\section{Methodology}


\begin{figure}[h]
  \centering
  \includegraphics[width=14cm]{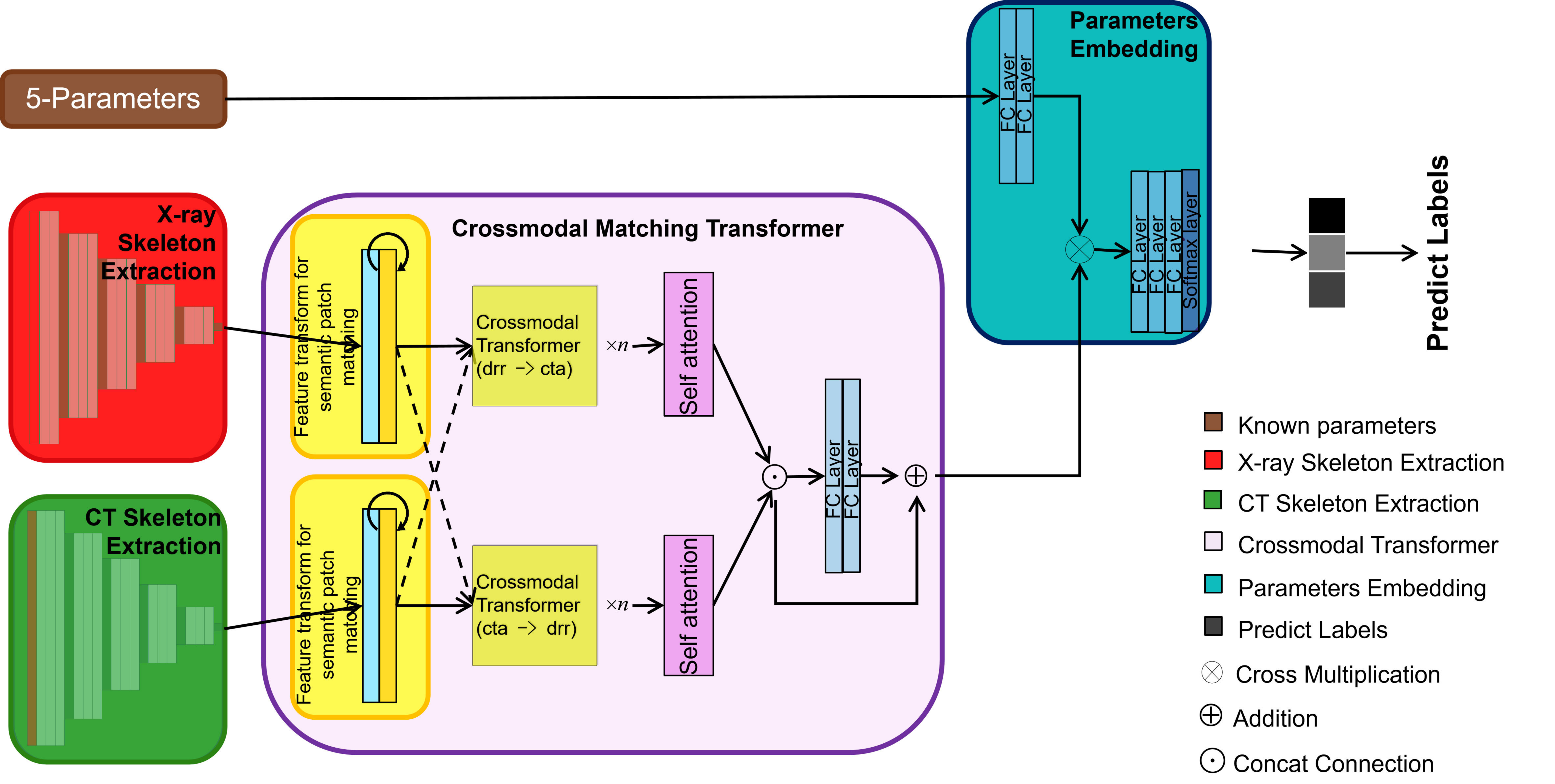}
  \caption{The entire network structure of Crossmodal Matching Transformer for intervention in TEVAR}
  \label{entire_net}
\end{figure}

\subsection{Feature extraction}

\noindent Our approach starts with extracting skeletons from 2D X-ray and 3D CT images. In particular, we train two UNet models for 2D skeleton segmentation and 3D skeleton segmentation tasks, as UNet has been successfully applied in various image segmentation tasks \cite{ronneberger2015u,cciccek20163d}. After they are well trained, their encoders can consequently provide strong skeleton-related representations, and thus we take them as the input for downstream tasks.

The 2D encoder is made up of five blocks that contains three 2D convolution units( a $3 \times 3$ convolution layer, a batch normalization layer (BN) and a ReLU activation) and a $3 \times 3$ max pooling layer with stride of two. Meanwhile, the 3D encoder is also made up of five blocks, where the first block has two 3D convolution units ( a $3 \times 3 \times 3$ convolution layer, a batch normalization layer (BN) and a ReLU activation ) and a $3 \times 3 \times 3$  max pooling with stride of two while Each of the rest four blocks containing three 3D convolution units and a 3D max pooling. In this paper, the output of the 2D and 3D encoder are $512$-channel 2D feature maps and $64$-channel 3D feature maps, representing the skeleton locations and appearance information of the 2D and 3D input.


\subsection{Deep Feature Matching}

\noindent After the feature extraction of 2D and 3D encoders way, the 2D and 3D skeleton representations are directly combined for registration rather than previous approaches \cite{miao2018dilated, liao2019multiview} which convert 3D image to 2D then registration.

Full network structure and more details please see the official article published in August 2021.

\subsection{Additional parameters}

Five known parameters: the angle of L-arm rotation($\alpha$), the angle of pivot rotation($\beta$), the angle of C-arm rotation($\gamma$), the distance from source to patient($SSD$), the distance from source to detector($SID$), and the coordinate of isocenter in pre-interventional CT data. 

\begin{figure}[h]
  \centering
  \includegraphics[width=8cm]{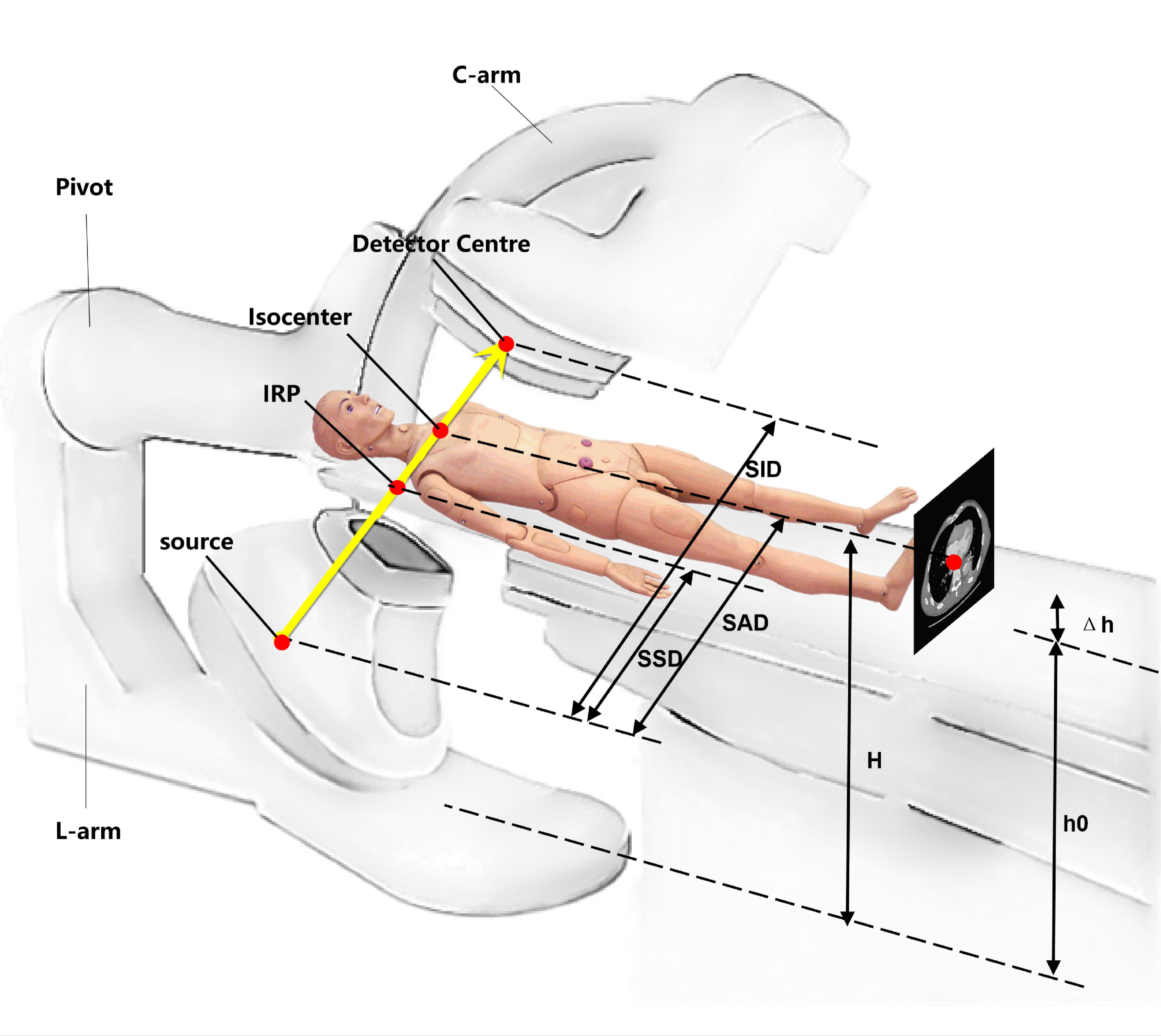}
  \caption{C-arm}
  
  \label{Fig.para}
\end{figure}

Full network structure and more details please see the official article published in August 2021.

\section{Experiment Results and Analysis}

\subsection{Implementation details}

\subsubsection{Dataset}
\label{Sec.para}
\begin{figure}[h]
  \centering
  \includegraphics[width=10cm]{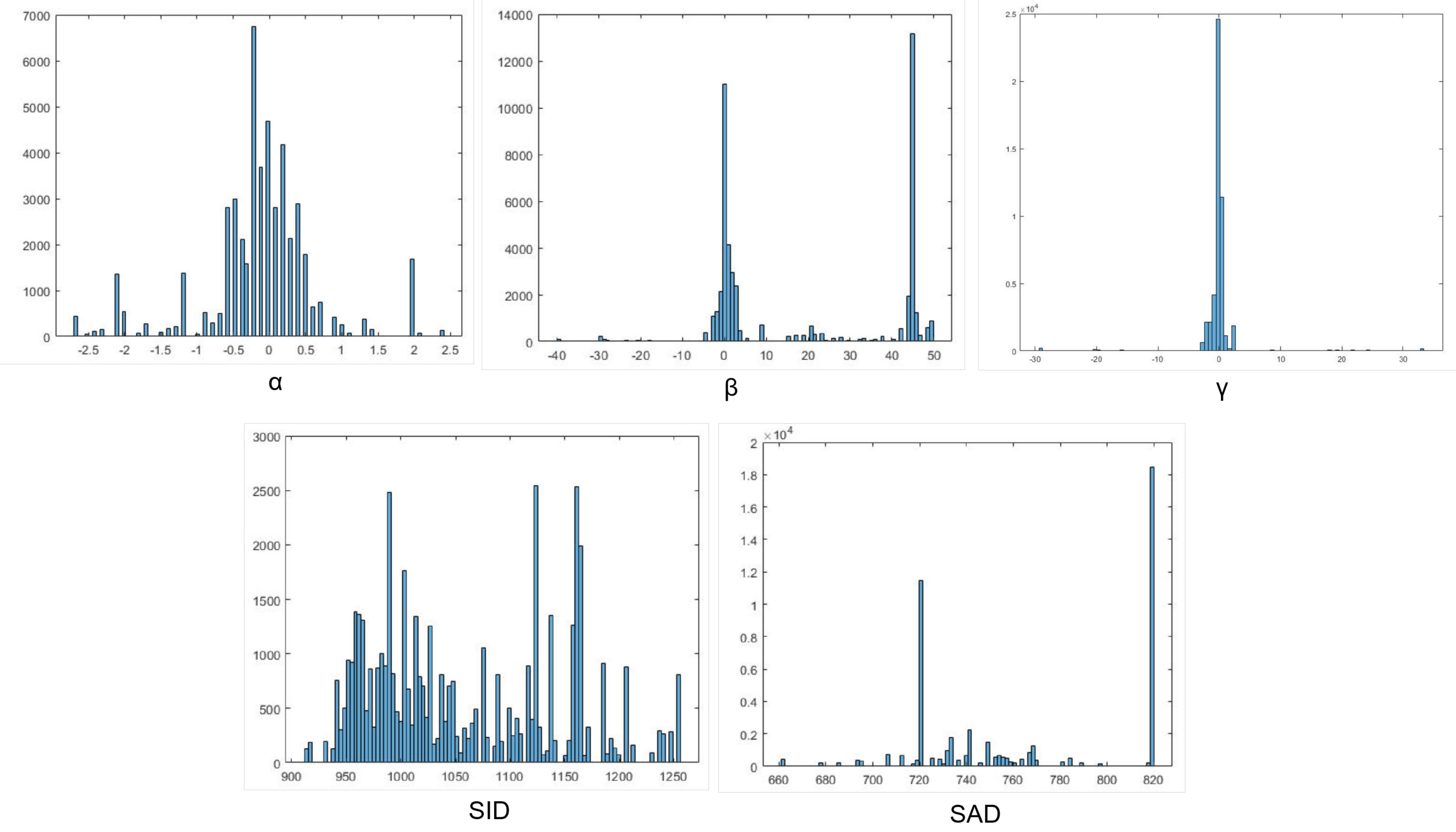}
  \caption{Distribution of location-related parameters available from imaging equipment in TEVAR}
  \label{fig.para_5}
\end{figure}

Pre-interventional CT images of 1162 patients suffering thoracic aortic aneurysm or B aortic dissection were collected for training of 3D segmentation network. Among them, 1000 were used for training and 162 for validation. 
A total of 49176 frames of X-ray examination parameters were counted in 162 patients. Their distribution is shown in Fig.\ref{fig.para_5}. 
According to the distribution range of the above parameters and random sampling of the position coordinates of isocenter in CT space, the clinical projection transformation space is constructed. 

Virtual intra-interventional X-rays are generated by ray casting to CT data with known clinical spatial transformation parameters. 
The input to 2D segmentation network is a $352\times512$ image with 1 channels, which is cut out from the upper part of virtual X-ray for removing the part of the diaphragm with larger motion. 
The corresponding skeleton label is segmented from CT data using the threshold method. 
After that, the $873$ groups pre-interventional CT images are tracing to generate $87300$ virtual X-rays according to the range shown in table.\ref{tab.para_space} and distribution shown in Fig.\ref{fig.para_5}. where $d$ represents the scanning diameter of CT data.

\begin{table}[h]
  \caption{Related Parameters}
  \label{tab.para_space}
  \begin{tabular}{rcccccccc}  
    \toprule
    &$\alpha$ &$\beta $ &$\gamma$ &$SID(mm)$  &$SSD(mm)$ \\  
    \midrule
    lower &$\ang{-2}$ &$\ang{0}$ &$\ang{-10}$ &850  &720\\
    upper &$\ang{2}$ &$\ang{50}$ &$\ang{10}$ &1250  &820\\  
  \bottomrule
  \end{tabular}
\end{table}

\subsubsection{Training details}

We implement the proposed approach under the Pytorch framework with GPU acceleration. 
3D segmentation network was trained by 1000 sets of CT data and their skeletal labels and verified by 162 sets. 
And 2D skeleton segmentation network was trained by 543 groups virtual X-ray with a total of 543000 images and validated by 162 groups with a total of 97000 images. 
These virtual X-ray is generated according to the parameter distribution of clinical X-ray, as described in Section.\ref{Sec.para}. 

Full network structure and more details please see the official article published in August 2021.

\subsection{Feasibility analysis of rigid registration to aortic arch}
\label{Sec.feasibility}
During interventional surgery, the doctor estimated the location of the branches of the aorta according to the position of the bone and the intraoperative guide wire under X-ray. 
In order to verify the rigid registration of bone can locate the aortic arch and its branches, we measured the displacement of vertebral bone, rib, guide wire and proximal stent in 309 groups intra-interventional X-ray. 
Because the vessel can not be clearly developed under X-ray, we use the displacement of the proximal stent post-intenventional to approximate estimates the displacement of the aortic arch and its branches. 
As a validation of the feasibility of our method.

\begin{table}[h]
  \caption{The displacement of intra-interventional vertebral bone, rib, guide wire and post-interventional proximal stent }
  \label{tab.displacement error}
  \begin{tabular}{rcc}  
    \toprule
    Object detected & Maximum$(mm)$ & Mean$(mm)$ \\  
    \midrule
    Guide wire & 9.52 & 3.76\\  
    proximal stent & 8.14 & 2.95\\
    rib $\ang{0}$ & 7.78 & {\bfseries 0.78}\\
    rib $\ang{45}$& 2.34 & 2.61\\
    vertebral bone $\ang{0}$ & 0.78 & {\bfseries 0.31} \\
    vertebral bone $\ang{45}$ & 1.67 & {\bfseries 0.91} \\
  \bottomrule
\end{tabular}
\end{table}

Since the average dynamic displacement error of branches vessels on aortic arch is about $2.95 mm$, the opening diameter of the left subclavian artery is about 10 mm and the respect of proximal anchoring area over 15 mm, we have good reason to think that rigid registration can achieve the initial localization of the aorta in TEVAR.

\subsection{Registration experiment}

We compare our approach to a learning-based(CNN\cite{miao2018dilated}) and two optimization-based methods. 
Cross-correlation(Opt-CC) and normalized mutual information(Opt-NMI) are used to measure the similarity  respectively and Powell algorithm is used to optimize similarity iterativly. 
In addition,  we compare the performance of the optimization-based aprroach using two learning-based aprroaches as initializers(Denoted as CNN+Opt and Cross-Transformer+Opt). 

The position of the aortic branch on DSA image at same time is taken as the gold standard, and the registration error of these approaches(mean target registration error, mTRE) is measured. The gold standard data is untrained. 
The gross failure rate(GFR) is defined as the percentage of th tested cases with a TRE greater than $10mm$. And the average time cost of registration(Reg.time) and out-plane rotation error(Rot.error) additionally demonstrate of the superiority of the approach. 

For detailed details of the genuine articles published in August.

Experimental results show that the accuracy and speed of registration are improved to some extent by using learning-based approach to initialize the optimization-based approach. 
The average displacement error of $4.7mm$ is considered to be of clinical significance in TEVAR operation. 
Although this method is still not very robust in the comprehensive performance of registration, it is carried out without additional positioning equipment or markers. 

\section{Conclusion}

In this paper, we propose a novel 3D-2D medical image registration approach. The main novelty of this paper is the proposed feature transform method that allows the generated feature maps retain the initial pose estimation of any 3D CT and corresponding 2D X-ray, without any other positioning items. The excellent experimental results shows that the proposed approach is suitable for basic clinical intervention usage and provides a new direction for intra-interventional registration.

Although deep learning has been widely used in the field of computer vision, its imprecision limits its application in the medical image registration. 
The approach of directly regression parameters is constrained by network performance, whose performance is not accurate enough to classify parameters. 
Pixel level accuracy is the most important problem to be solved in the field of registration. 
Other areas, such as image recognition, require translation rotation invariance, which is strictly prohibited in medical image registration. 
Therefore, the reference network structure should be more rigorous. 



\bibliographystyle{ACM-Reference-Format}
\bibliography{sample-base}

\appendix

\end{document}